      \documentstyle[preprint,aps]{revtex}

\begin{document}
\draft
\begin{title}
SOFT TURBULENCE
IN THE ATMOSPHERIC BOUNDARY LAYER
\end{title}
\author{\bf Imre M. J\'anosi}
\begin{instit}
Department of Atomic Physics, E\"otv\"os University
Budapest, Puskin u. 5-7, H-1088, Hungary
Fax: + 36-1-266-0206
E-mail: JANOSI\char'100rs1.comphys.uni-duisburg.de
\end{instit}
\moreauthors{\bf G\'abor Vattay}
\begin{instit}
Department of Solid State Physics, E\"otv\"os University
Budapest, M\'uzeum krt. 6-8, H-1088, Hungary
Fax: + 36-1-266-7509
E-mail: VATTAY\char'100NBIVAX.NBI.DK
\end{instit}

\begin{abstract}

In this work we compare the spectral properties
of the daily medium temperature fluctuations with the experimental results of
the Chicago Group, in which the local temperature fluctuations were measured
in a helium cell.  The results suggest that the dynamics of the daily
temperature fluctuations is determined by the soft turbulent state of the
atmospheric boundary layer, which state is significantly different from low
dimensional chaos.

\end{abstract}

\vspace{0.6cm}

There is a widespread interest in forecasting and modeling of weather.
Although all the basic mechanisms that govern the dynamics of the atmosphere
have been well known since quite a long time, the detailed understanding and
adequate characterization of the fluctuations of various statistical
quantities of the lower atmospheric boundary layer is still not complete. For
example, there were several attempts to describe the dynamics of daily medium
temperature fluctuations. A typical viewpoint is that the underlying mechanism
is fully stochastic in nature and can be considered as an autoregressive
process.  A completely different viewpont suggests that the apparent
irregularities may be attributed to a deterministic chaotic behavior, although
serious doubts have arisen on the existence of low dimensional chaos in the
long time behavior of the atmosphere (climatic attractor), as well as in the
processes over very short time scales.

The present investigation is based on temperature measurements by the Hungarian
Meteorological Service performed at twenty different meteorological
stations covering the area of Hungary for the period 1.1.1951 -- 31.12.1989.
The detailed analysis has been published in Ref. [1]. We produced the
fluctuation time series by substracting the deterministic part from the daily
medium temperature data. The histogram of the fluctuation amplitudes has a
pronounced Gaussian distribution\cite{mi}.
The power density spectrum
can be obtained by well established methods,
Figure 1 shows a typical result.

\newpage

\vspace*{8cm}

\figure{\it Unnormalized power density spectrum of the temperature fluctuations
measured in Szombathely. The frequency unit is 1/day. The solid
line is the fit given by  Eq. (1), the dashed line illustrates the $1/f^2$
behavior on a restricted frequency range.}

\bigskip

 All of the spectra of the time series measured at different meteorological
stations can be fitted by the
function
$$ P(f) = P_s \exp \left[ - \left( {f\over f_s}\right)^\beta \right]\quad,
\eqno(1)$$
where $P_s=222\pm 16$, $f_s=0.017\pm 0.005$, $\beta =0.54 \pm 0.03 $, and
the deviations indicate slight meteorological station dependence.
Surprisingly, this form of the fitting function (1)  with the same exponent
$\beta$, and the Gaussian fluctuation-histogram are exactly the same as
found in the helium experiment\cite{ch4} in the soft turbulent region (see
later).
We note that the high
frequency part of the power spectrum may be fitted by a power law on a
very restricted frequency range (approximately half decade) with an exponent
$\sim -2$, which inspired the Markovian stochastic models.

The shape of the power spectrum and the Gaussian fluctuation distribution do
not rule out theoretically the existence of a low dimensional meteorological
attractor. However, the embedding process\cite{gras} applied to our data did
not show any saturation of the correlation dimensions. This observation
completely agrees with the measurement of Talkner, Weber, and
Roser\cite{suiss}, in which they could not find a weather attractor with a
dimension less than 10 from a longer temperature time series .

Our conclusions are based on the detailed investigations of the Chicago Group,
in which they have measured the local temperature fluctuations in gaseous
helium at very wide parameter ranges\cite{ch4,ch1}.
The control parameter in these experiments is the
dimensionless Rayleigh number\cite{ch1} $R$.
Four different domains in $R$ were observed.

\vspace*{9.5cm}

\figure{\it Nusselt number versus Rayleigh number measured in a helium cell,
the domains for the various transitions are defined. [F. Heslot, B. Castaing,
and A. Libchaber, Phys. Rev. {\bf A 36}, 5870 (1987).]}

\bigskip

In the first domain, the onset of convection,
 the onset of an oscillatory instability,
and the onset of a chaotic state could be easily
identified\cite{ch1}. The correlation dimension $D$ of the chaotic
state\cite{gras} was determined\cite{ch1}, which was $D\approx 2$ at the onset
of chaos, and increased rapidly reaching a value of $D=4$ at $R=2\times 10^5$.
The second domain from $R=2.5\times 10^5$ to $R=5\times 10^5$ was a transition
region, where the coherence function\cite{ch1} between the bolometers decreased
rapidly and finally disappeared.
The third domain is known as the soft turbulent
regime, and the transition from soft to the so called hard turbulence occurs
at $R\approx 10^8$, which seems to be universal value.
The main differences between hard and soft turbulence are the following:
The probability distribution function of the local temperature fluctuations in
hard turbulence is exponential while for soft turbulence is Gaussian.
Moreover, the power spectrum of the local temperature fluctuations is
streched-exponential in soft turbulence [Eq. (1)], while the low-frequency
range of the
power spectra clearly exhibits a power law behavior with an exponent $-7/5$ in
hard turbulence\cite{ch4}.

We have performed a detailed comparison of the two measurement  in Ref.
[1].
The conclusions are the following:
\begin{enumerate}
\item One can estimate the characteristic height $L$ of the convecting
atmospheric
layer using the fitted cutoff frequency  $f_s$ of Eq. (1) and known
parameters, such as the thermal diffusivity and the Prandtl number of the air.
The result ($L\approx 30-500$ m) is in agreement with the accepted values of
the thickness of the air layer influenced by the daily cycle of temperature
change.
\item The atmospheric boundary layer is usually considered as a
layer of infinite aspect ratio.  We think that this problem might be resolved
by observations, which suggest that the vertical and horizontal sizes of the
medium scale ($\sim 100-500$ m) convective eddies are approximately
equal.
\item The spectral properties and the probability distribution of the
daily medium temperature fluctuations suggest that  the atmospheric boundary
layer exhibits a soft turbulent thermal convection.
\item As the soft turbulent state occurs after several transitions from
chaotic state, connected with the increase of the number of effective
degrees of freedom,  it is unlikely that the typical atmospheric dynamics
exhibits low dimensional chaos.
\end{enumerate}

This work has been supported by the Hungarian Scientific Research Foundation
(OTKA)  under Grant No. 521. and 2091.

\end{document}